\documentclass[12pt]{iopart}
\input{epsf.sty}
\def\3j#1#2#3#4#5#6{ \left(
  \begin{array}{ccc}
  #1 & #2 & #3\\
  #4 & #5 & #6
  \end{array}
  \right)}
\def\6j#1#2#3#4#5#6{  \left\{
  \begin{array}{ccc}
  #1 & #2 & #3 \\
  #4 & #5 & #6
  \end{array}
  \right\} }
\def\9j#1#2#3#4#5#6#7#8#9{ \left\{
  \begin{array}{ccc}
  #1 & #2 & #3\\
  #4 & #5 & #6 \\
  #7 & #8 & #9
  \end{array}
  \right\}}
\begin{document}

\title[Alkali doublets perturbed by helium]{Collisional broadening of
alkali doublets by helium perturbers}

\author{D F T Mullamphy$^1$, G Peach$^2$, V Venturi$^1$\footnote{Present
address: Centre for Vascular Research, University of New South Wales,
Sydney 2052, Australia}, I B Whittingham$^1$ and S J Gibson$^1$}

\address{$^1$ School of Mathematics, Physics and Information Technology,
James Cook University, Townsville 4811, Australia}

\address{$^2$ Department of Physics and Astronomy, University College
London WC1E~6BT, UK}

\ead{ian.whittingham@jcu.edu.au}

\begin{abstract}
We report results for the Lorentzian profiles of the Li~I, Na~I and K~I
doublets and the Na I subordinate doublet  broadened by helium perturbers for
temperatures up to 3000~K. They have been obtained from a fully
quantum-mechanical  close-coupling description of the colliding atoms, the
Baranger theory of line shapes and new \textit{ab initio} potentials for
the alkali-helium interaction. For all lines except the 769.9 nm K I line,
the temperature dependence of the widths over the range
$70 \leq T \leq 3000$~K is accurately represented by the power law form
$w=aT^b$ with  $0.37 < b < 0.43$. The 769.9 nm K I line has this form for
$500 \leq T \leq 3000$~K  with $b$ having the higher value of 0.49.
Although the shifts have a more complex temperature dependence, they all
have the general feature of increasing with temperature above $T \sim 500$~K
apart from the 769.9 K I line whose shift decreases with temperature.
\end{abstract}


\submitto{\jpb}
\maketitle

\section{Introduction}

Accurate pressure broadened profiles of alkali resonance doublets
perturbed by H$_{2}$ and He are of crucial importance for the modelling
of atmospheres of late M, L and T type brown dwarfs and for generating
their synthetic spectra in the region 0.6 - 1.1 $\mu$m.  The dominant lines
are the Na I 589.0/589.6 nm and K I 766.5/769.9 nm doublets
\cite{Pav2000,Bur2001} which occur in the middle of a spectral region
where background absorption is particularly small so that both the line
centres and wings stand out.
There can also be significant contributions from less abundant alkalis
such as Li, Rb and Cs, and from subordinate doublets such as
Na I 818.3/819.5 nm.

The profiles of the strongly broadened alkali doublets have been
the subject of several recent studies using models primarily designed
to describe the non-Lorentzian far wings of the profiles.
Burrows and Volobuyev \cite{Bur2003} used a quasistatic unified
Franck-Condon semiclassical model
to study the Na I and K I doublets whereas a unified line shape
semiclassical model
has been used by Allard \etal \cite{All2003} for the Na I and
K I doublets and, more
recently, for the Rb I and Cs I doublets \cite{All2006}. Alternatively,
quantum mechanical single-channel models based upon
bound-free and free-free transitions in the transient
molecules formed during collision have been used to calculate the
emission and absorption spectra for the wings of the
Li I \cite{Mason,Zhu2005}, Na I \cite{Mason,Zhu2006}
and K I \cite{Zhu2006} resonance lines in the presence of He perturbers.
Despite these studies, highly accurate calculations of the central
Lorentzian cores are still
needed in order to estimate the effects of dust in brown
dwarf atmospheres.

We report results for the widths and shifts of the Lorentzian profiles
over the temperature
range $70 \leq T \leq 3000$~K of  the Li I, Na I and K I doublets
and the Na I subordinate doublet broadened by
helium perturbers.  The calculations extend our previous study
\cite{Leo2000} of pressure broadening of the Na I doublets by He at
laboratory temperatures up to 500 K and are based on a fully
quantum-mechanical close-coupling description of the colliding atoms and
the Baranger \cite{Bar58} theory of line shapes.
As input for our calculations
we have computed new \textit{ab initio} potentials for the alkali-helium
interaction. Preliminary results for the widths of two Na I lines
for temperatures up to 2000 K have been given in \cite{Peach2005} and for the
widths of the Na I and K I doublets in \cite{Peach2006}\footnote{Note
that the results reported in \cite{Peach2006} are for full half-widths and
not half half-widths as stated}.

\section{Interatomic potentials}

The adiabatic molecular potentials ${}^{2S+1}V_{\Lambda}(R)$ for the
X$^{*}$--He system (where X = Li, Na or K) have been obtained by
using a three-body model in which the alkali X$^{*}$ is treated as
a X$^{+}$ ion plus an active electron and the perturbing He atom is
represented by a polarizable atomic core.  Model potentials are
used to represent the electron-atom and electron-atomic ion
interactions and the basic methods adopted for obtaining these
model potentials are discussed by \cite{Peach82}.
These potentials reproduce not only the correct energies of the bound
states of X$^{*}$ but also generate wavefunctions for both the bound and
scattering states of X$+e^{-}$ and He$+e^{-}$ that contain the correct
number of nodes. The model potentials also support unphysical bound
states corresponding to the presence of closed shells in the X${^+}$
ion and in He. This effect is taken into account in the calculation of
the molecular potential by including the unphysical states in the
atomic basis used in the diagonalization of the Hamiltonian for the
three-body model.

The most important recent change from the earlier approach is that
the core-core interaction is itself calculated directly using a
three-body model composed of X$^+$ and He$^+$ cores plus one electron.
A more detailed discussion of this development is given for the Li-He
case by \cite{Behmenburg}.  Further numerical refinements have been
incorporated into the computer program since then and new calculations
carried out for the three alkali-helium systems considered in our study.

\begin{figure}
\begin{center}
\epsfsize=0.8\textwidth\epsfbox{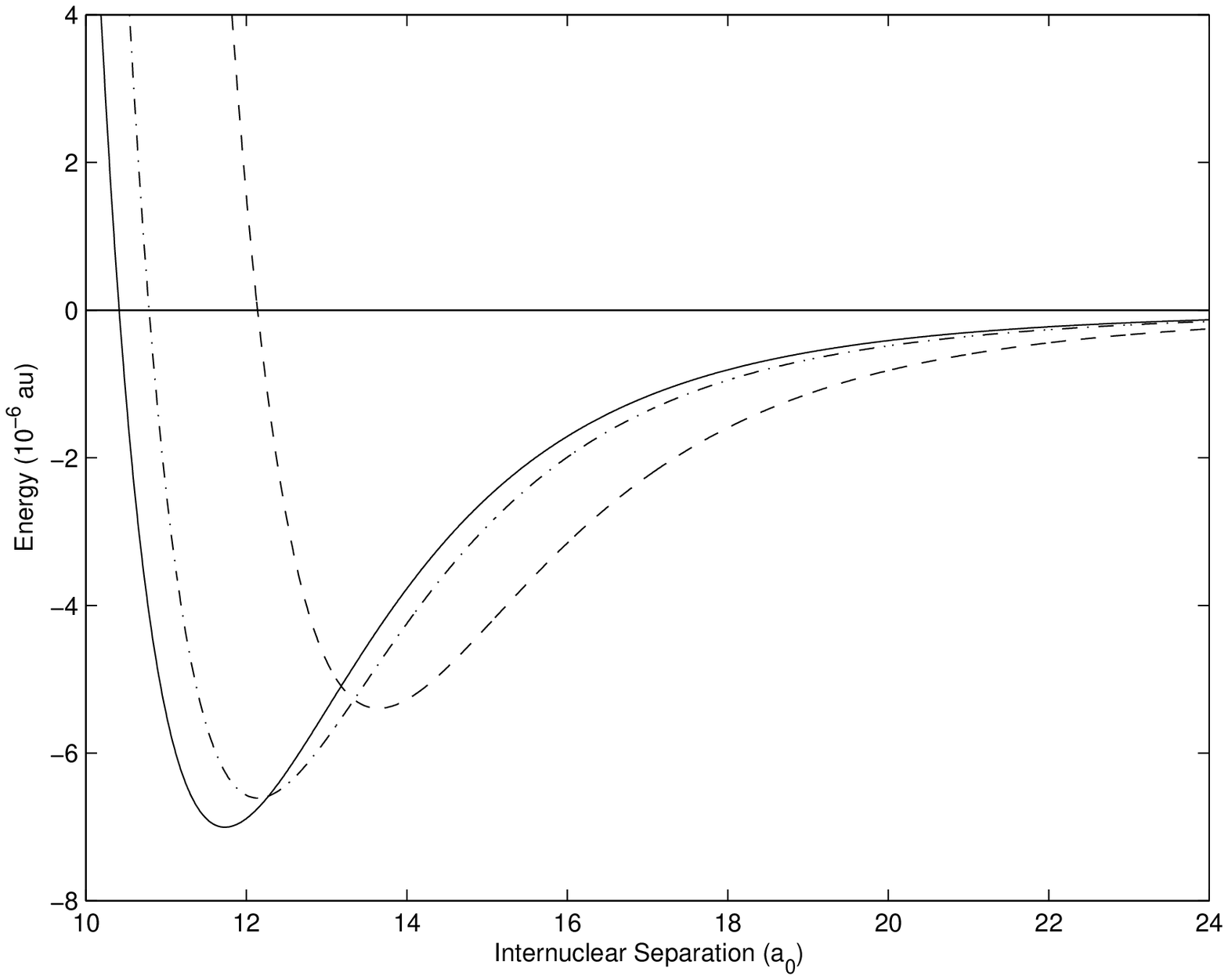}
\end{center}
\caption{\label{ns}Adiabatic ns~$^2V_{\Sigma}$ molecular potentials:
Li-He (\full), Na-He (\chain) and K-He (\broken). }
\end{figure}

\begin{figure}
\begin{center}
\epsfsize=0.8\textwidth\epsfbox{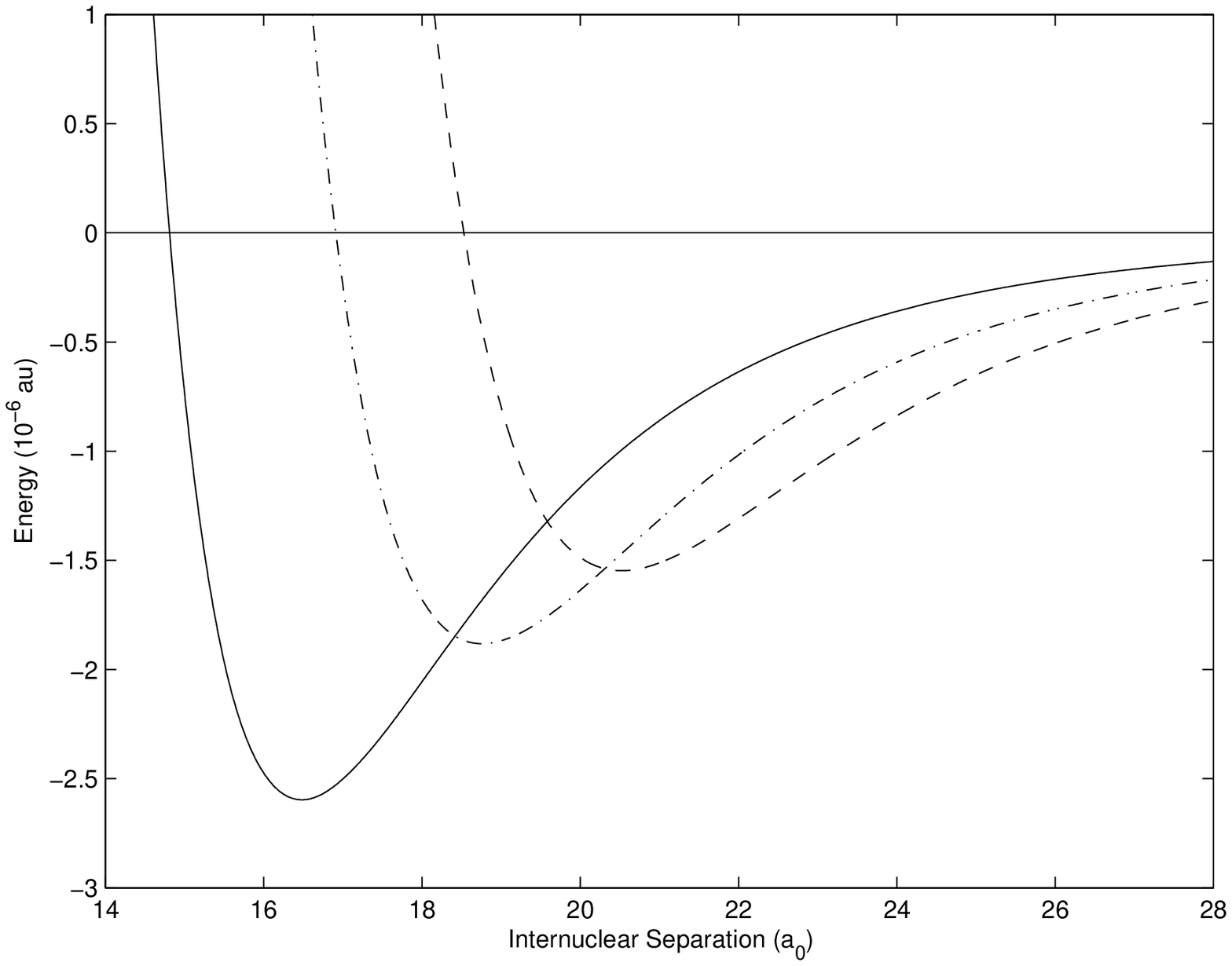}
\end{center}
\caption{\label{np0}Adiabatic np~$^2V_{\Sigma}$ molecular potentials:
Li-He (\full), Na-He (\chain) and K-He (\broken).}
\end{figure}

\begin{figure}
\begin{center}
\epsfsize=0.8\textwidth\epsfbox{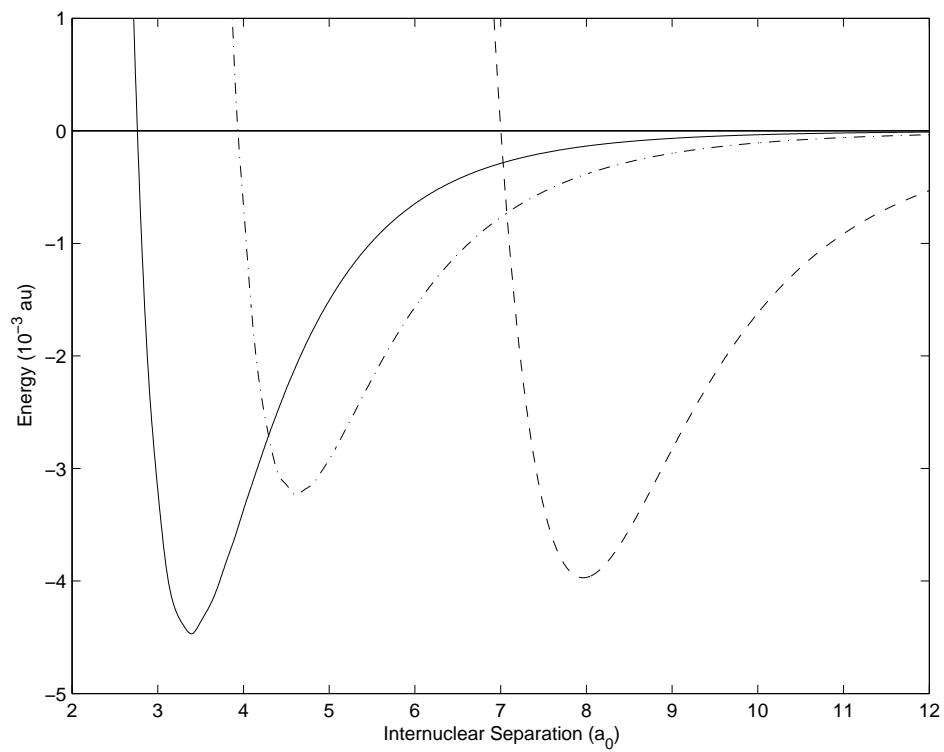}
\end{center}
\caption{\label{np1}Adiabatic np~$^2V_{\Pi}$ molecular potential:
Li-He (\full ), Na-He (\chain) and K-He (\broken).
Note that the Na-He and K-He potentials have been scaled up by factors
of 5 and 20 respectively.}
\end{figure}

\begin{figure}
\begin{center}
\epsfsize=0.8\textwidth\epsfbox{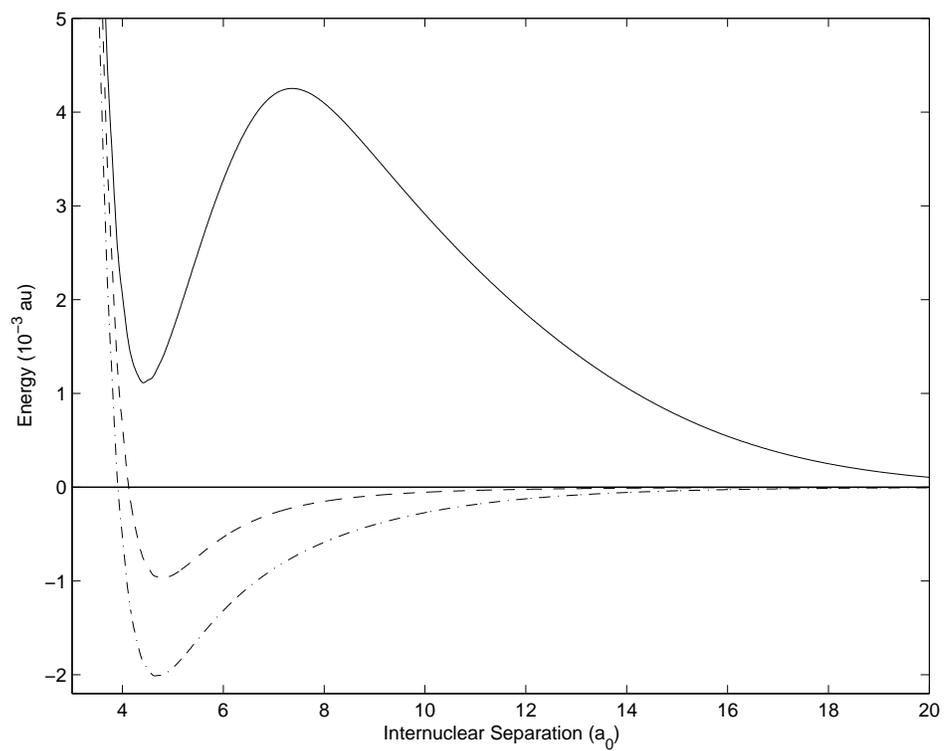}
\end{center}
\caption{\label{Napd}Adiabatic potential curves for Na-He:
3d~$^2V_{\Sigma }$ (\full ),
3d~$^2V_{\Pi}$ (\chain ) and 3d $^2V_{\Delta }$ (\broken ).}
\end{figure}

The ns $^2V_{\Sigma}$, np $^2V_{\Sigma}$ and np $^2V_{\Pi}$ molecular
potentials for Li-He $(n=2)$, Na-He $(n=3)$ and K-He $(n=4)$ are shown
in figures \ref{ns}, \ref{np0} and \ref{np1} respectively.  The Na-He
3d $^2V_{\Sigma,\Pi,\Delta}$ potentials are shown in figure \ref{Napd}.
The minima in the potentials used and their positions are given in tables
\ref{Lipots}, \ref{Napots} and \ref{Kpots} and compared with other
data where they are available.  New \textit{ab initio} potentials for K-He
have recently been obtained \cite{pot14} but no details are given.
  The overall agreement with other work is
very satisfactory and the three-body model can be expected to give
good results at medium and large interatomic separations.  However the
positions of the minima in the np $^2V_{\Pi}$ potentials occur at short
range and are largely determined by the behaviour of the core-core
interaction.  For the Li-He and Na-He cases the agreement with theory
and experiment for the np $^2V_{\Pi}$ states is generally good, but the
result for the K-He 4p $^2V_{\Pi}$ state requires some comment.  Because
the 3p virtual state supported by the K$^+ + e^-$ model potential has a
higher energy than that of He(1s$^2$), the core-core potential, K$^+$-He,
is unreliable at small separations.  Methods for resolving this
difficulty are still being investigated, but since the behaviour of
line widths and shifts depends largely on the difference between upper
and lower state potentials they are not particularly sensitive to this
problem.

\section{Spectral line profile}

The width and shift of each spectral line have been
calculated using the quantum-mechanical impact theory of
Baranger \cite{Bar58} for non-overlapping spectral lines in which the
profile of each isolated line is a Lorentzian.
The  orbital, spin and total electronic angular momentum
operators for the alkali atom are $\mathbf{L}$, $\mathbf{S}$ and
$\mathbf{j}=\mathbf{L}+\mathbf{S}$ respectively and radiation is emitted
by the alkali atom as it  undergoes a transition between initial
and final  states $|n_{i}L_{i}S_{i}j_{i}\rangle $ and
$|n_{f}L_{f}S_{f}j_{f}\rangle $,  where $n$ denotes the  principal quantum
number.
The half half-width $w$ and shift $d$ at temperature $T$ of the
Lorentzian profile is given by \cite{Leo95}
\begin{equation}
\label{nahe1}
w + \rmi d = N\; \int_{0}^{\infty}\;f(E)\;S(E)\;\rmd E
\end{equation}
where $f(E)$ is the normalized Maxwellian perturber energy distribution
\begin{equation}
\label{nahe2}
f(E) = 2 \pi (\pi k_{{\rm B}} T)^{-3/2}\:\sqrt{E}\;\exp[-E/(k_{{\rm B}}T)],
\end{equation}
$N$ is the perturber number density and
\begin{eqnarray}
\label{nahe3}
\fl
  S(E) = \frac{\hbar^{2}\pi}{M^{2}} \sqrt{\frac{M}{2E}}\;
  \sum_{l,l^{\prime}}\;\sum_{J_{i},J_{f}}
(2J_{i}+1)(2J_{f}+1)(-1)^{l+l^{\prime}}
\left\{{J_{f}\;J_{i}\;1\atop j_{i}\;j_{f}\;l}\right\}
\left\{{J_{f}\;J_{i}\;1\atop j_{i}\;j_{f}\;l^{\prime}}\right\} \nonumber \\
\times   [\delta_{l,l^{\prime}}-
\langle j_{i}\;l^{\prime}\;J_{i}|S|j_{i}\;l\;J_{i}\rangle
\langle j_{f}\;l^{\prime}\;J_{f}|S|j_{f}\;l\;J_{f}\rangle ^{*}]
\end{eqnarray}
describes the effects of collisions on the two states forming the
spectral line. Here $M$ is the reduced mass of the emitter-perturber system,
$l$ and $l^{\prime}$ are quantum numbers corresponding to the relative
emitter-perturber angular momentum $\mathbf{L}_{R}$ before and after the
collision and $\mathbf{J}=\mathbf{L}_{R}+\mathbf{j}$
is the total angular momentum of the emitter-perturber system.
The scattering matrix element in the coupled  $|j\;l\;J\rangle $
representation is  $\langle j^{\prime}\;l^{\prime}\;J|S|j\;l\;J\rangle $
where we have suppressed the quantum numbers $(n,L,S)$ for convenience
and $\left\{{a\;b\;c\atop f\;g\;h}\right\}$
is the $6-j$  symbol \cite{Edmonds74}.

The scattering matrix elements in (\ref{nahe3}) are determined from the
asymptotic behaviour of the radial functions $G^{J}_{jl}(R)$ for each
scattering channel $(j,l,J)$.  These functions satisfy the coupled equations
\cite{Leo95}
\begin{equation}
\label{nahe4}
  [\frac{\partial^{2}}{\partial R^{2}}
-\frac{l(l+1)}{R^{2}}+ k_{j}^{2}]
G^{J}_{jl,j^{\prime\prime}l^{\prime\prime}}(R)=
\frac{2M}{\hbar^{2}}\sum_{j^{\prime},l^{\prime}}
\;V^{J}_{jl,j^{\prime}l^{\prime}}(R)
\;G^{J}_{j^{\prime}l^{\prime},j^{\prime\prime}
l^{\prime\prime}}(R),
\end{equation}
where $(j^{\prime \prime}, l^{\prime \prime})$ labels
the linearly independent solutions of (\ref{nahe4}) and
the Born-Oppenheimer coupling terms have been neglected. The parameter
\begin{equation}
\label{nahe5}
k_{j}^{2}= 2M[E-(E^{(L,S)}+\varepsilon_{j}(\infty ))]/\hbar^{2}
\end{equation}
is positive for open scattering channels and negative for closed channels.
Here $E$ is the total energy of the emitter-perturber system,
$E^{(L,S)}$ is the energy of the state of the  separated atoms to which
the molecular state dissociates
adiabatically and the fine structure parameter $\varepsilon_{j}(R)$
has been assumed to have its asymptotic value $\varepsilon_{j}(\infty )$.
The interaction potential matrix elements
$V^{J}_{jl,j^{\prime}l^{\prime}}(R)$ are, for the ns $^{2}\mathrm{S}_{j}$
level
\begin{equation}
\label{nahe7a}
V_{jl,j^{\prime}l^{\prime}}(R) = \delta_{j,j^{\prime}}\delta_{l,l^{\prime}}
\;^{2}V_{\Sigma}(R),
\end{equation}
for the np $^{2}\mathrm{P}_{j}$ levels,
\begin{equation}
\label{nahe7}
V_{jl,j^{\prime}l^{\prime}}(R) = \delta_{j,j^{\prime}}\delta_{l,l^{\prime}}
\;^{2}V_{\Pi}(R)+C^{(1)}_{jl,j^{\prime}l^{\prime}}[^{2}V_{\Sigma}(R)-
\;^{2}V_{\Pi}(R)]
\end{equation}
and, for the nd $^{2}\mathrm{D}_{j}$ levels\footnote{We correct here the
phase factor given in \cite{Leo2000} for the $^{2}\mathrm{D}_{j}$ levels.}
\begin{eqnarray}
\label{nahe9}
\fl  V_{jl,j^{\prime}l^{\prime}}(R) =  \delta_{j,j^{\prime}}
\delta_{l,l^{\prime}}\;^{2}V_{\Delta}(R)+
(-1)^{j^{\prime}-j}[1+(-1)^{l^{\prime}+l}] B_{jl,j^{\prime}l^{\prime}}
[^{2}V_{\Pi}(R)-\;^{2}V_{\Delta}(R)]   \nonumber \\
+C^{(2)}_{jl,j^{\prime}l^{\prime}}[^{2}V_{\Sigma}(R)-\;^{2}V_{\Delta}(R)]
\end{eqnarray}
where the coefficients
\begin{eqnarray}
\label{nahe8}
\fl  C^{(n)}_{jl,j^{\prime}l^{\prime}}=\sum_{\Omega}\;(-1)^{j^{\prime}-j}
C(J\;j\;l;-\Omega\;\Omega\;0)
C(J\;j^{\prime}\;l^{\prime};-\Omega\;\Omega\;0)   \nonumber \\
\times C(n\;\case{1}{2}\;j;0\;\Omega\;\Omega)
C(n\;\case{1}{2}\;j^{\prime};0\;\Omega\;\Omega)
\end{eqnarray}
and
\begin{eqnarray}
\label{nahe10}
\fl  B_{jl,j^{\prime}l^{\prime}}=\sum_{\Omega}\;
C(J\;j\;l;-\Omega\;\Omega\;0)
C(J\;j^{\prime}\;l^{\prime};-\Omega\;\Omega\;0) \nonumber \\
\times C(2\;\case{1}{2}\;j;1\;\Omega-1\;\Omega)
C(2\;\case{1}{2}\;j^{\prime};1\;\Omega-1\;
\Omega)
\end{eqnarray}
are symmetric under $(j,l)\leftrightarrow (j^{\prime},l^{\prime})$.
Here $C(j_{1}j_{2}j_{3};\Omega_{1}\Omega_{2}\Omega_{3})$ is the
Clebsch-Gordan coefficient,
$\Omega $ denotes the projection of an angular momentum onto the
internuclear axis $\bi{R}$ and
${}^{2S+1}V_{\Lambda}(R)$, where $\Lambda \equiv |\Omega_{L}|$, are
the adiabatic molecular potentials.

The equations (\ref{nahe4}) decouple into two sets of opposite parity
$(-1)^{J\pm 1/2}$. Thus the  ns $^{2}\mathrm{S}_{j}$,
np $^{2}\mathrm{P}_{j}$ and nd $^{2}\mathrm{D}_{j}$ states give rise to
one, three and five coupled differential equations respectively for each
parity. These equations were solved using a modified
version \cite{Leo95} of the
R-Matrix method of Baluja {\it et al} \cite{Bal82} and the solutions
fitted to free-field boundary conditions to extract the scattering matrix
elements.

\section{Results and discussion}

Calculations have been completed for the
Li I doublet
2p ${}^{2}\mathrm{P}_{3/2,1/2} \rightarrow $ 2s ${}^{2}\mathrm{S}_{1/2}$,
the Na I doublet
3p ${}^{2}\mathrm{P}_{3/2,1/2} \rightarrow $ 3s ${}^{2}\mathrm{S}_{1/2}$,
the K I doublet
4p ${}^{2}\mathrm{P}_{3/2,1/2} \rightarrow $ 4s ${}^{2}\mathrm{S}_{1/2}$
and the Na I subordinate "doublet"
3d ${}^{2}\mathrm{D}_{3/2} \rightarrow $ 3p ${}^{2}\mathrm{P}_{1/2}$ and
3d ${}^{2}\mathrm{D}_{5/2,3/2} \rightarrow $ 3p ${}^{2}\mathrm{P}_{3/2}$.
The main
computational issues associated with extending our previous
calculations up to the higher temperatures (3000 K) of
astrophysical interest are convergence of the sum over partial waves
in (\ref{nahe3}) and of the integration (\ref{nahe1}) over
perturber energies. The present calculations required about
500 partial waves for most of the energy nodes but up to 1000
partial waves were needed at the highest energy nodes which corresponded
to temperatures extending to 5000 K.

The $R$-matrix propagation was commenced using the arbitrary choice
$R=I$ at a distance $R_{\mathrm{min}}$ well within the classical
turning point so that the solutions were independent of $R_{\mathrm{min}}$.
Typically $R_{\mathrm{min}}$ ranged from $2a_{0}$ for small $l$ to
$25a_{0}$ for large $l$. The distance $R_{\mathrm{max}}$ at which the
solutions were matched to the free-field solutions was typically
$300a_{0}-400a_{0}$. This yielded $S$-matrix elements accurate to at
least six significant figures and, after summing over partial waves,
resulted in six figure accuracy for $\Re S(E)$ and four for $\Im S(E)$.
As the major part of the $R$-matrix method is energy independent, equations
(\ref{nahe4}) are solved at each value of $l$ for the
entire set of energy nodes.
The large number of partial waves needed might suggest
that a semiclassical treatment would be more appropriate. However the
present quantal calculations did not require an inordinate amount of
computer time  (typically a few hours) and are free of the uncertainties
endemic in semiclassical calculations of determining the appropriate
lower cutoff on impact parameters in order to exclude perturber paths
entering the classically forbidden regions of the interatomic
interaction  \cite{Leo95}.

\begin{figure}
\begin{center}
\epsfsize=0.8\textwidth\epsfbox{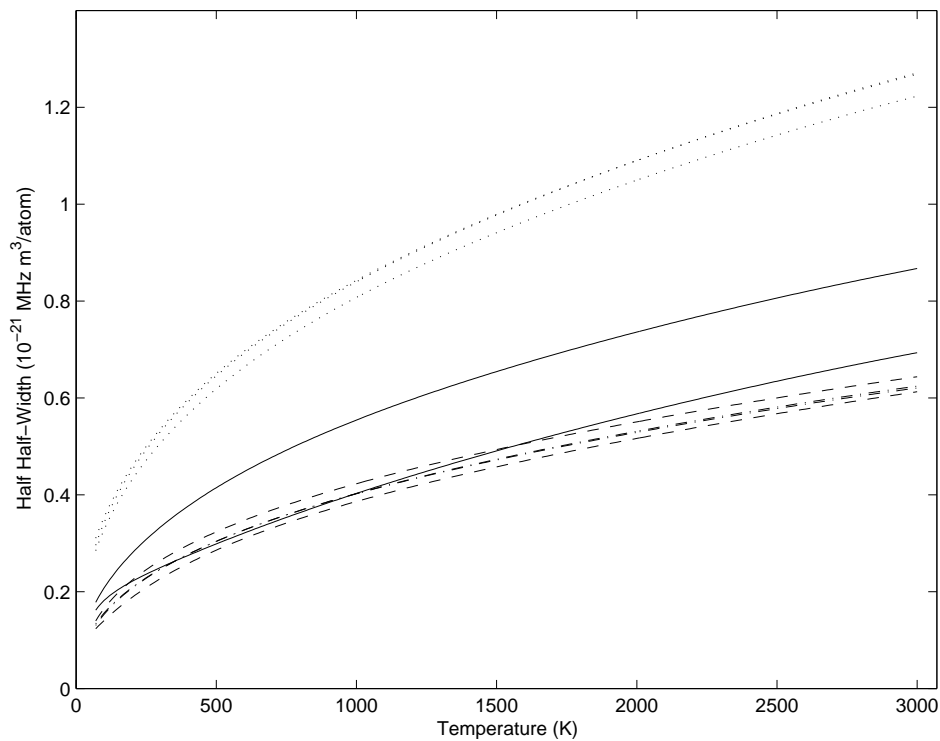}
\end{center}
\caption{\label{fwidths}Temperature dependence of half half-widths
(in units of  $10^{-21}$ MHz m$^{3}$/atom) of alkali lines perturbed by He.
The doublets np $^2\mathrm{P}_{j}$ -- ns $^2\mathrm{S}_{1/2}$ shown
are Li (\chain), Na (\broken) and K (\full) where the upper (lower) curve
corresponds to $j=3/2 (j=1/2)$. The Na lines
3d $^2\mathrm{D}_{j_{i}}$ -- 3p $^2\mathrm{P}_{j_{f}}$ are shown
(\dotted) with the upper curve corresponding to $(j_{i},j_{f})=(3/2,3/2)$
and $(3/2,1/2)$ unresolved and the lower curve to $(5/2,3/2)$.
Note that, on the scale shown, the members of the Li doublet cannot be
resolved.}
\end{figure}

\begin{figure}
\begin{center}
\epsfsize=0.8\textwidth\epsfbox{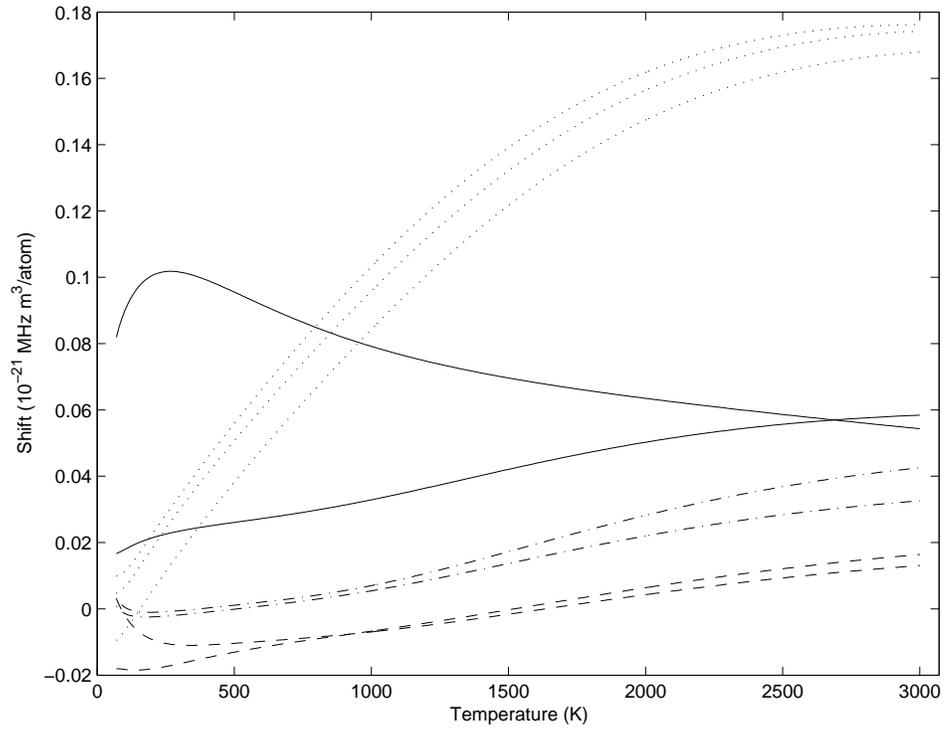}
\end{center}
\caption{\label{fshifts}Temperature dependence of shifts (in units of
$10^{-21}$ MHz m$^{3}$/atom) of alkali lines perturbed by He.
The doublets np $^2\mathrm{P}_{j}$ -- ns $^2\mathrm{S}_{1/2}$ shown
are Li (\chain), Na (\broken) and K (\full) where the upper (lower) curve
at the lower temperatures corresponds to $j=1/2 (j=3/2)$. The Na lines
3d $^2\mathrm{D}_{j_{i}}$ -- 3p $^2\mathrm{P}_{j_{f}}$ lines are shown
(\dotted) with $(j_{i},j_{f})=(5/2,3/2)$ (upper curve), $(3/2,3/2)$
(middle curve) and $(3/2,1/2)$ (lower curve).}
\end{figure}

The temperature dependence of the computed half half-widths and shifts
are shown in
figures \ref{fwidths} and \ref{fshifts} respectively.
The widths and shifts of all lines were calculated at
10 K intervals and in all cases  were found to be smooth functions
of temperature.  The widths are accurately represented to
three significant figures by  the power law form
\begin{equation}
\label{nahe11}
w(T) = aT^{b}
\end{equation}
where the fit parameters $a$ and $b$ are given in table \ref{widths}.  Also
shown are the values of $b$ obtained by \cite{Lwin77} using a
semiclassical model which does not resolve the members of each doublet.
Our results agree closely with this early calculation except for
the anomalous 4p $^{2}\mathrm{P}_{1/2}$ -- 4s $^{2}\mathrm{S}_{1/2}$
potassium transition where our value is significantly higher.
For all lines the temperature dependence is
stronger than the $T^{0.3}$ behaviour obtained from a classical
treatment using a pure van der Waals $V(R) = -C_{6}/R^{6}$ interaction.

Although the shifts have a more complex temperature dependence, they all
have the general feature of increasing with temperature above $T \sim 500$~K
apart from the
4p $^{2}\mathrm{P}_{1/2}$ -- 4s $^{2}\mathrm{S}_{1/2}$ potassium transition
which is again anomalous as its shift decreases with temperature.

The effects of fine structure on the calculated widths and shifts arises
from the $6-j$ symbols in (\ref{nahe3}) and the $j$-dependence of the
$S$-matrix due to the potential matrix elements
$V_{jl,j^{\prime}l^{\prime}}(R)$ and the fine-structure splittings
$\varepsilon_{j}(\infty)$ appearing in (\ref{nahe4}). However, as
the zero energy for the calculation of any level ${}^{2S+1}\mathrm{L}_{j}$
is set at the energy $E_{0}=E^{(L,S)}+\varepsilon_{j}(\infty)$ of that
level and the splittings are
negligible for all but the lowest energy nodes, the role of the
actual splittings
is relatively unimportant. This is evident in the results for the Na lines
3d ${}^{2}\mathrm{D}_{5/2,3/2} \rightarrow $ 3p ${}^{2}\mathrm{P}_{3/2}$
where the widths and shifts are significantly different even though
$\varepsilon_{3/2}(\infty)=\varepsilon_{5/2}(\infty)$.

The behaviour of the anomalous potassium transition warrants some comment.
The calculation for the potassium 4p $^{2}\mathrm{P}_{1/2}$ level is
different from that of all other levels considered in the present study
in that there are closed channels at the lowest energy nodes and the
wells of the molecular potentials are very shallow
(depths $E_{d} < 10^{-4}$ au), smaller than the fine structure splitting
$\varepsilon =2.68 \times 10^{-4}$ au of the level.
Consequently the channels only become open at energies above
$\varepsilon $ and therefore at energies above $E_{d}$.

We compare our calculated widths and shifts with measurement in
table \ref{wcomps} and table \ref{scomps} respectively.
We have not included all the experimental  width data for Na as
the present calculations for the Na doublets closely reproduce our
earlier results \cite{Leo2000} for $T \leq 500$ K and a detailed
comparison with the numerous existing theoretical and experimental studies
was reported in that paper. In general the widths are in good agreement
for all alkalis  although only one measurement is
for $T > 700$ K. The situation regarding the shifts is less
clear. The predicted shift for the
3d $^2\mathrm{D}_{3/2}$ -- 3p $^2\mathrm{P}_{1/2}$ Na I line agrees very
closely with the measurement of \cite{Beh90} and the results for
the K I doublet have the same sign and relative magnitude as the
measurements. However the predicted shifts for the Li I and Na I
doublets have the opposite sign to that of the measured shifts.
The shifts are quite sensitive to the precise details of the potentials
as they are produced by a balance between the effects of the long-range
attractive potential and the short-range repulsive potential. In
particular, for a given energy they are sensitive to where the repulsive
wall is located. Consequently the disagreement with experiment for the
Li I and Na I doublets suggests that the repulsive region of the Li
potential may need to be slightly suppressed and that of the Na potential
enhanced.

\ack
The authors would like to thank Tim Beams for his helpful advice on
computing issues.

\Tables

\Table{\label{Lipots} Positions $R_{d}\,(a_{0})$ and minimum energies
$V_{d}$ (au) of the Li-He adiabatic molecular potentials. Also given are
the $C_{6}$ (au) coefficients for the asymptotic forms of the potentials.}
\br
\centre{2}{2s $^2V_{\Sigma}$} & \centre{2}{2p $^2V_{\Sigma}$} &
\centre{2}{2p $^2V_{\Pi}$}  \\ 
&&&&& \\
\crule{2} & \crule{2} & \crule{2} \\
$R_{d}$ & $V_{d}$ & $R_{d}$ & $V_{d}$ & $R_{d}$ & $V_{d}$  \\
\mr
11.73 & $-7.005 \times 10^{-6}{\,}^{\rm a}$ &
16.48 & $-2.596 \times 10^{-6}{\,}^{\rm a}$ &
3.39 & $-4.470\times 10^{-3}{\,}^{\rm a}$ \\
11.553 & $-6.787\times 10^{-6}{\,}^{\rm b}$ &
17.027 & $-1.511 \times 10^{-6}{\,}^{\rm c}$ &
3.395 & $-4.55 \times 10^{-3}{\,}^{\rm d}$ \\
11.56 & $-6.79 \times 10^{-6}{\,}^{\rm e}$ &
18.0 & $-1.4 \times 10^{-6}{\,}^{\rm f}$ &
3.44 & $-5.65 \times 10^{-3}{\,}^{\rm g}$ \\
12.2 & $-9.1\times 10^{-6}{\,}^{\rm k}$ &
17.0 & $-3.1\times 10^{-6}{\,}^{\rm h}$ &
3.44 & $-4.670\times 10^{-3}{\,}^{\rm h}$ \\
12.0 & $-4.6 \times 10^{-6}{\,}^{\rm f}$ &
&& 3.44 & $-4.07 \times 10^{-3}{\,}^{\rm i}$ \\
11.385 & $-1.18 \times 10^{-5}{\,}^{\rm h}$ &
&& 3.40 & $-4.638 \times 10^{-3}{\,}^{\rm j}$ \\
11.6 & $-6.47 \times 10^{-6}{\,}^{\rm l}$ &
&& 3.60 & $-3.14\times 10^{-3}{\,}^{\rm k}$ \\
&&&& 3.42 & $-3.95 \times 10^{-3}{\,}^{\rm f}$ \\
&&&& 3.37 & $-4.647 \times 10^{-3}{\,}^{\rm l}$ \\
&&&&& \\
$C_{6} =22.57{\,}^{\rm a}$ & & $C_{6}=50.75{\,}^{\rm a}$ & &
$C_{6}=28.31{\,}^{\rm a}$ & \\
\br
\end{tabular}
\item[]$^{\rm a}$ Theory, this work.
\item[]$^{\rm b}$ Theory \cite{pot5,pot9,pot11,pot15}
\item[]$^{\rm c}$ Theory \cite{pot1,pot9,pot11}
\item[]$^{\rm d}$ Theory \cite{Behmenburg,pot9,pot11}
\item[]$^{\rm e}$ Theory \cite{pot5}
\item[]$^{\rm f}$ Theory \cite{pot16}
\item[]$^{\rm g}$ Theory \cite{pot8}
\item[]$^{\rm h}$ Theory \cite{pot2,pot16}
\item[]$^{\rm i}$ Theory \cite{pot13}
\item[]$^{\rm j}$ Theory \cite{pot7}
\item[]$^{\rm k}$ Theory \cite{pot15}
\item[]$^{\rm l}$ Experiment \cite{pot10}
\end{indented}
\end{table}

\Table{\label{Napots} Positions $R_{d}\,(a_{0})$ and minimum energies
$V_{d}$ (au) of the Na-He adiabatic molecular potentials. Also shown is the
position $R_{p}\,(a_{0})$ and energy $V_{p}$ (au) of the peak in the
3d $^2V_{\Sigma}$ potential together with the
$C_{6}$ (au) coefficients for the asymptotic forms of the potentials.}
\br
\centre{2}{3s $^2V_{\Sigma}$} & \centre{2}{3p $^2V_{\Sigma}$} &
\centre{2}{3p $^2V_{\Pi}$}  \\
&&&&& \\
\crule{2} & \crule{2} & \crule{2} \\
$R_{d}$ & $V_{d}$ & $R_{d}$ & $V_{d}$ & $R_{d}$ & $V_{d}$  \\
\mr
12.14 & $-6.610 \times 10^{-6}{\,}^{\rm a}$ &
18.77 & $-1.881 \times 10^{-6}{\,}^{\rm a}$ &
4.61 & $-1.617\times 10^{-3}{\,}^{\rm a}$ \\
12.0 & $-1.3\times 10^{-5}{\,}^{\rm b}$ &
&& 4.22 & $-3.01 \times 10^{-3}{\,}^{\rm b}$ \\
12.2 & $-5.56 \times 10^{-6}{\,}^{\rm f}$ &
&& 4.35 & $-3.33 \times 10^{-3}{\,}^{\rm c}$ \\
&&&& 4.35 & $-2.33\times 10^{-3}{\,}^{\rm d}$ \\
&&&& 4.42 & $-2.34 \times 10^{-3}{\,}^{\rm e}$ \\
&&&& 4.35 & $-2.04 \times 10^{-3}{\,}^{\rm f}$ \\
&&&& 4.58 & $-1.4\times 10^{-3}{\,}^{\rm g}$ \\
&&&& 4.4 & $-2.19 \times 10^{-3}{\,}^{\rm h}$ \\
&&&&& \\
$C_{6} =26.33{\,}^{\rm a}$ & & $C_{6}=77.02{\,}^{\rm a}$ & &
$C_{6}=43.72{\,}^{\rm a}$ & \\
\br
\centre{2}{3d $^2V_{\Sigma}$} & \centre{2}{3d $^2V_{\Pi}$} &
\centre{2}{3d $^2V_{\Delta}$}  \\
&&&&& \\
\crule{2} & \crule{2} & \crule{2} \\
$R_{d}$ & $V_{d}$ & $R_{d}$ & $V_{d}$ & $R_{d}$ & $V_{d}$  \\
\mr
4.41 & $1.115 \times 10^{-3}{\,}^{\rm a}$ &
4.64 & $-2.010 \times 10^{-3}{\,}^{\rm a}$ &
4.80 & $-9.717 \times 10^{-4}{\,}^{\rm a}$ \\
33.08 & $-2.167 \times 10^{-7}{\,}^{\rm a}$ & &&&& \\
\crule{2} & & && \\
$R_{p}$ & $V_{p}$ & & & & \\
\crule{2} &&&& \\
7.37 & $4.252 \times 10^{-3}{\,}^{\rm a}$ &&&&& \\
&&&&& \\
$C_{6} =218.7{\,}^{\rm a}$ & & $C_{6}=194.3{\,}^{\rm a}$ & &
$C_{6}=121.2{\,}^{\rm a}$ & \\
\br
\end{tabular}
\item[]$^{\rm a}$ Theory, this work.
\item[]$^{\rm b}$ Theory \cite{pot4}
\item[]$^{\rm c}$ Theory \cite{pot8}
\item[]$^{\rm d}$ Theory \cite{pot2}
\item[]$^{\rm e}$ Theory \cite{pot13}
\item[]$^{\rm f}$ Theory \cite{pot7}
\item[]$^{\rm g}$ Theory \cite{pot3}
\item[]$^{\rm h}$ Experiment \cite{pot6}
\end{indented}
\end{table}

\Table{\label{Kpots} Positions $R_{d}\,(a_{0})$ and minimum energies
$V_{d}$ (au) of the K-He adiabatic molecular potentials. Also shown are the
$C_{6}$ (au) coefficients for the asymptotic forms of the potentials.}
\br
\centre{2}{4s $^2V_{\Sigma}$} & \centre{2}{4p $^2V_{\Sigma}$} &
\centre{2}{4p $^2V_{\Pi}$}  \\
&&&&& \\
\crule{2} & \crule{2} & \crule{2} \\
$R_{d}$ & $V_{d}$ & $R_{d}$ & $V_{d}$ & $R_{d}$ & $V_{d}$  \\
\mr
13.65 & $-5.395 \times 10^{-6}{\,}^{\rm a}$ &
20.52 & $-1.547 \times 10^{-6}{\,}^{\rm a}$ &
7.96 & $-1.986\times 10^{-4}{\,}^{\rm a}$ \\
13.9 & $-3.5\times 10^{-6}{\,}^{\rm c}$ &
&& 5.29 & $-2.19 \times 10^{-3}{\,}^{\rm b}$ \\
11.5 & $-2.85 \times 10^{-5}{\,}^{\rm e}$ &
&& 5.29 & $-1.03 \times 10^{-3}{\,}^{\rm c}$ \\
&&&& 5.30 & $-1.12\times 10^{-3}{\,}^{\rm d}$ \\
&&&& 5.4 & $-7.7 \times 10^{-4}{\,}^{\rm e}$ \\
&&&&& \\
$C_{6} =42.24{\,}^{\rm a}$ & & $C_{6}=105.6{\,}^{\rm a}$ & &
$C_{6}=61.96{\,}^{\rm a}$ & \\
\br
\end{tabular}
\item[]$^{\rm a}$ Theory, this work.
\item[]$^{\rm b}$ Theory \cite{pot8}
\item[]$^{\rm c}$ Theory \cite{pot7}
\item[]$^{\rm d}$ Theory \cite{pot2}
\item[]$^{\rm e}$ Theory \cite{pot17}
\end{indented}
\end{table}

\Table{\label{widths}Parameters $a$ and $b$ for the fit $w = aT^b$
to the temperature dependence  of the pressure broadened Lorentzian
half half-widths (in units of $10^{-21}$ MHz m$^{3}$/atom) of various alkali
doublets due to He perturbers. The results apply to the temperature range
70 -- 3000 K except for the 4p $^{2}\mathrm{P}_{1/2}$ --
4s $^{2}\mathrm{S}_{1/2}$ potassium transition where the fit applies to
the range 500 -- 3000 K. Numbers in parentheses denote the uncertainty in
the last significant figure from the fitting procedure.}
\br
Element & Transition & $\lambda $ (nm) & $a$ & $b$ &
$b_{\mathrm{LML}}^{\rm a}$ \\
\mr
Li & 2p $^2\mathrm{P}_{1/2}$ -- 2s $^2\mathrm{S}_{1/2}$ & 670.78 &
0.02533(5)  & 0.3998(2) & 0.39 \\

Li & 2p $^2\mathrm{P}_{3/2}$ -- 2s $^2\mathrm{S}_{1/2}$ & 670.79 &
0.02461(5) & 0.4042(3) & \\

Na & 3p $^2\mathrm{P}_{1/2}$ -- 3s $^2\mathrm{S}_{1/2}$ & 589.59 &
0.02011(6) & 0.4270(4) & 0.40 \\

Na & 3p $^2\mathrm{P}_{3/2}$ -- 3s $^2\mathrm{S}_{1/2}$ & 589.00 &
  0.02918(8)  &  0.3866(4) &  \\

Na & 3d $^2\mathrm{D}_{3/2}$ -- 3p $^2\mathrm{P}_{1/2}$ & 818.33 &
  0.06075(8) &  0.3799(2) & \\

Na & 3d $^2\mathrm{D}_{3/2}$ -- 3p $^2\mathrm{P}_{3/2}$ & 819.48 &
  0.06369(5) & 0.3737(1) & \\

Na & 3d $^2\mathrm{D}_{5/2}$ -- 3p $^2\mathrm{P}_{3/2}$ & 819.48 &
  0.05755(9) & 0.3820(2) & \\

K & 4p $^2\mathrm{P}_{1/2}$ -- 4s $^2\mathrm{S}_{1/2}$  & 769.90 &
   0.01385(8) &  0.4886(8) & 0.39 \\

K & 4p $^2\mathrm{P}_{3/2} $ -- 4s $^2\mathrm{S}_{1/2}$ & 766.49 &
  0.03173(7) & 0.4136(4) &  \\
\br
\end{tabular}
\item[]$^{\rm a}$ Theoretical results \cite{Lwin77} from
semiclassical model which does not resolve the doublet.
\end{indented}
\end{table}

\Table{\label{wcomps} Lorentzian half half-widths
(in units of $10^{-21}$MHz m$^{3}$/atom) for alkali doublets
pressure broadened by He perturbers.}
\br
Element & Transition &  $T$(K) & Theory & Experiment \\
\mr
Li & 2p $^2\mathrm{P}_{1/2}$ -- 2s $^2\mathrm{S}_{1/2}$ & 600 &
0.3274  & $0.302 ^{\rm a}$ \\

Li & 2p $^2\mathrm{P}_{3/2}$ -- 2s $^2\mathrm{S}_{1/2}$ & 600 &
0.3274 & $0.302 ^{\rm a}$ \\

Li & 2p $^2\mathrm{P}_{1/2,3/2}$ -- 2s $^2\mathrm{S}_{1/2}$ & 673 &
0.3431  & $0.33 \pm 0.17^{\rm b}$ \\

Na & 3p $^2\mathrm{P}_{1/2}$ -- 3s $^2\mathrm{S}_{1/2}$ & 450 &
0.2731 & $0.260\pm 0.003 ^{\rm c}$ \\
&&&& $0.242 \pm 0.016 ^{\rm d}$ \\

Na & 3p $^2\mathrm{P}_{3/2}$ -- 3s $^2\mathrm{S}_{1/2}$ & 450 &
0.3096   & $0.318 \pm 0.006 ^{\rm c}$   \\
&&&& $ 0.258 \pm 0.016 ^{\rm d} $ \\

Na & 3p $^2\mathrm{P}_{1/2}$ -- 3s $^2\mathrm{S}_{1/2}$ & 480 &
0.2807 & $0.302\pm 0.016 ^{\rm e}$ \\

Na & 3p $^2\mathrm{P}_{3/2}$ -- 3s $^2\mathrm{S}_{1/2}$ & 480 &
0.3174   & $0.349 \pm 0.032 ^{\rm e}$   \\

Na & 3p $^2\mathrm{P}_{1/2,3/2}$ -- 3s $^2\mathrm{S}_{1/2}$ & 2450 &
0.567 & $0.560\pm 0.064 ^{\rm g}$ \\
&&&& $0.726 \pm 0.064 ^{\rm h}$ \\

Na & 3d $^2\mathrm{D}_{3/2}$ -- 3p $^2\mathrm{P}_{1/2}$ & 470 &
0.6290 &  $0.645 \pm 0.02 ^{\rm f}$  \\

K & 4p $^2\mathrm{P}_{1/2}$ -- 4s $^2\mathrm{S}_{1/2}$  & 410 &
0.2773 & $0.247 ^{\rm a}$ \\

K & 4p $^2\mathrm{P}_{3/2} $ -- 4s $^2\mathrm{S}_{1/2}$ & 410 &
0.3808 &  $0.328 ^{\rm a }$  \\
\br
\end{tabular}
\item[]$^{\rm a}$ \cite{Lwin77}
\item[]$^{\rm b}$ \cite{Gall75}
\item[]$^{\rm c}$ \cite{Kiel80}
\item[]$^{\rm d}$ \cite{Del73}
\item[]$^{\rm e}$ \cite{McCar76}
\item[]$^{\rm f}$ \cite{Beh90}
\item[]$^{\rm g}$ \cite{BehKoh64}
\item[]$^{\rm h}$ \cite{Beh64}
\end{indented}
\end{table}

\Table{\label{scomps} Lorentzian shifts
(in units of $10^{-21}$MHz m$^{3}$/atom) for alkali doublets
pressure broadened by He perturbers.}
\br
Element & Transition &  $T$(K) & Theory & Experiment \\
\mr
Li & 2p $^2\mathrm{P}_{1/2}$ -- 2s $^2\mathrm{S}_{1/2}$ & 630 &
0.0023  & \-0.045$ ^{\rm a}$ \\

Li & 2p $^2\mathrm{P}_{3/2}$ -- 2s $^2\mathrm{S}_{1/2}$ & 630 &
0.0011 & \-0.018$ ^{\rm a}$ \\

Na & 3p $^2\mathrm{P}_{1/2}$ -- 3s $^2\mathrm{S}_{1/2}$ & 450 &
\-0.0107 & $0.048\pm 0.014 ^{\rm b}$ \\

Na & 3p $^2\mathrm{P}_{3/2}$ -- 3s $^2\mathrm{S}_{1/2}$ & 450 &
\-0.0139   & $0.018 \pm 0.014 ^{\rm b}$   \\

Na & 3p $^2\mathrm{P}_{1/2}$ -- 3s $^2\mathrm{S}_{1/2}$ & 480 &
\-0.0105 & $0.006\pm 0.009 ^{\rm c}$ \\

Na & 3p $^2\mathrm{P}_{3/2}$ -- 3s $^2\mathrm{S}_{1/2}$ & 480 &
\-0.0134 & $0.011 \pm 0.011 ^{\rm c}$ \\

Na & 3d $^2\mathrm{D}_{3/2}$ -- 3p $^2\mathrm{P}_{1/2}$ & 470 &
0.0352 & $0.034 \pm 0.003 ^{\rm d}$ \\

K & 4p $^2\mathrm{P}_{1/2}$ -- 4s $^2\mathrm{S}_{1/2}$  & 410 &
0.0988 & $0.072 \pm 0.006 ^{\rm e}$  \\

K & 4p $^2\mathrm{P}_{3/2} $ -- 4s $^2\mathrm{S}_{1/2}$ & 410 &
0.0250 & $0.038 \pm 0.003 ^{\rm e}$ \\

\br
\end{tabular}
\item[]$^{\rm a}$ \cite{Harris82}
\item[]$^{\rm b}$ \cite{Kiel80}
\item[]$^{\rm c}$ \cite{Lwin76}
\item[]$^{\rm d}$ \cite{Beh90}
\item[]$^{\rm e}$ \cite{Lwin77}

\end{indented}
\end{table}

\end{document}